\documentclass[aps,prb,twocolumn]{revtex4-1}
\usepackage{graphicx}

\begin{document}

\title{PBr$_3$ adsorption on a chlorinated Si(100) surface with mono- and bivacancies}

\author{T. V. Pavlova$^{1,2}$}
\author{V. M. Shevlyuga$^{1}$}
\email{pavlova@kapella.gpi.ru}
\affiliation{$^{1}$Prokhorov General Physics Institute of the Russian Academy of Sciences, Moscow, Russia}
\affiliation{$^{2}$HSE University, Moscow, Russia}

\begin{abstract}

For the most precise incorporation of single impurities in silicon, which is utilized to create quantum devices, a monolayer of adatoms on the Si(100) surface and a dopant-containing molecule are used. Here we studied the interaction of a phosphorus tribromide with a chlorine monolayer with mono- and bivacancies in a scanning tunneling microscope (STM) at 77 K. The combination of different halogens in the molecule and the adsorbate layer enabled unambiguous identification of the structures after PBr$_3$ dissociation on Si(100)-Cl. A Cl monolayer was exposed to PBr$_3$ in the STM chamber, which allows us to compare the same surface areas before and after PBr$_3$ adsorption. As a result of this comparison, we detected small changes in the chlorine layer and unraveled the molecular fragments filling mono- and bivacancies. Using density functional theory, we found that the phosphorus atom occupies a bridge position after dissociation of the PBr$_3$ molecule, which primarily bonds to silicon in Cl bivacancies. These findings provide insight into  the interaction of a dopant containing molecule with an adsorbate monolayer on Si(100) and can be applied to improve the process of single impurities incorporation into silicon.

\end{abstract}

\maketitle

\section{Introduction}

One of the challenges of silicon functionalization is the atomically precise positioning of impurities in a given configuration. Such accurate impurity insertion into silicon is required for the implementation of silicon quantum computing utilizing the nuclear spin of an impurity \cite{1998Kane}. The most precise insertion is achieved using STM (scanning tunneling microscope) lithography on a resist from a hydrogen monolayer on the Si(100) surface, followed by phosphine adsorption \cite{2001OBrien, 2003Schofield}. The current state of the art precision of phosphorus atom positioning on the surface is several Si lattice sites \cite{2012Fuechsle, 2022Wyrick}. Although this accuracy should in principle be sufficient to implement two-qubit operations \cite{2020Voisin, 2021Joecker}, enhancing the accuracy of P positioning is highly desirable for improving the quality of quantum computing. One of the ways to achieve higher precision could be by exploiting other chemical reactions on the surface.

To precisely introduce impurities into silicon, a hydrogen resist is currently utilized \cite{2012Fuechsle, 2019He, 2022Wang, 2022Kiczynski}; however, other adsorbates can also serve as resists. To serve as a resist, the adsorbate must first of all cover chemically active Si dangling bonds, protecting the surface from unwanted incorporation. Second, it should be possible to remove individual adatoms in the STM to create a mask on the Si(100) surface. Third, the adsorbate layer must be stable at the temperature used for STM lithography and subsequent molecule adsorption. Fourth, given that silicon epitaxy is the next step in device creation \cite{2003Schofield}, the desorption temperature of the adatoms should be low enough to avoid extra silicon heating, which results in diffusion of precisely arranged impurities. Alternatively, the resist atoms must segregate onto the surface during epitaxy, and in this case, it is desirable that the resist promotes the formation of crystalline silicon at the lowest temperatures by increasing Si atom mobility during epitaxy.

Phosphine is the impurity-containing molecule utilized in the fabrication of quantum devices \cite{2001OBrien, 2003Schofield, 2012Fuechsle, 2022Wyrick}, however other molecules can also be employed \cite{2020Stock, 2021DwyerACS, 2021Radue, 2023Lundgren} if the following conditions are met. First, to use a molecule to incorporate single impurity, it should contain only one impurity atom. Second, the molecule must form a bond with silicon atoms, but must not penetrate under the resist. Third, the other atoms of the molecule, except for the impurity, should be desorbed at a sufficiently low temperature or segregated on the surface during epitaxy to not remain as defects in the silicon crystal.

In this work, we chose a chlorine monolayer as a resist, which, as previously demonstrated \cite{2018Pavlova, 2019Dwyer, 2020Pavlova, 2021Pavlova, 2021Frederick, 2022Pavlova, 2019Pavlova, 2022Farzaneh}, fulfills all of the resist's requirements. As a molecule containing phosphorus and compatible with chlorine resist, we selected phosphorus tribromide \cite{2023Shevlyuga}. We took a molecule with a different halogen because bromine and chlorine are clearly distinguishable in the STM, allowing us to identify which halogen atoms appeared on the surface after the dissociation of the molecule. Using STM and density functional theory (DFT), we studied the adsorption of PBr$_3$ on Si(100) covered with a chlorine monolayer that contains mono- and bivacancies. Vacancies are created due to incomplete coverage of the surface with chlorine. It was essential that we carried out PBr$_3$ adsorption in the STM chamber so that we could observe changes on the surface after adsorption. After PBr$_3$ adsorption, we observed small modifications in the chlorine layer and identified structures with PBr$_3$ fragments in mono- and bivacancies.

\begin{figure*}[t]
 \includegraphics[width=\linewidth]{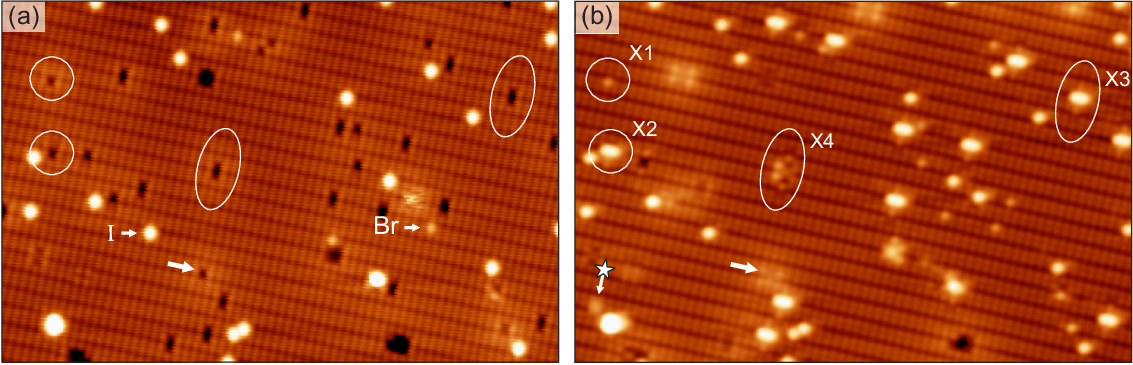}
\caption{\label{fig_big} Chlorinated Si(100)-2$\times$1 surface before and after exposure to PBr$_3$. (a) Empty state STM image (24.0$\times$15.4\,nm$^2$,  $U_s =3.5$\,V, I$_t$ = 2.0\,nA) of the initial Si(100)-2$\times$1 surface covered with an incomplete monolayer of chlorine with a predominance of mono- and bivacancies. The monovacancies are circled, the bivacancies are indicated by ellipses. The iodine and bromine atoms are also indicated in the STM image. (b) Empty state STM image (24.0$\times$15.4\,nm$^2$,  $U_s =2.8$\,V, I$_t$ = 2.0\,nA) of the same surface area after PBr$_3$ adsorption. The objects formed in the vacancies are labelled as X1--X4. The arrow indicates a monovacancy that has moved to an adjacent silicon atom. A new object formed on a defect-free area of the Cl monolayer is indicated by an arrow with an asterisk.}
\end{figure*}

\section{Methods}

\subsection{Experimental methods}

The experiments were conducted in an ultra-high vacuum (UHV) setup with a base pressure of 5$\times$10$^{-11}$\,Torr equipped with GPI CRYO (SigmaScan Ltd.) operating at 77\,K. To prepare the Si(100) samples (B-doped, 1\,$\Omega$\,cm), the wafers were outgassed at 870\,K for several days in UHV, then flash-annealed at 1470\,K. After the flash heating was turned off, Cl$_2$ was introduced at a sample temperature of 370--420\,K to prevent the formation of defects. Molecular chlorine was supplied directly to the sample surface by a fine leak valve for two seconds at a partial pressure in the chamber of $2 \cdot 10^{-9}$\,Torr. After chlorine adsorption, iodine and bromine also appeared on the Si(100) surface due to their presence in the inlet line used in previous experiments. Iodine and bromine concentrations were low enough that their presence on the surface had no impact with the experiment. Instead, the presence of iodine simplified the search for the area of the initial surface after PBr$_3$ adsorption since the iodine adatoms did not move whereas the chlorine layer had minor alterations. PBr$_3$ vapors were introduced into the STM chamber via a tube passing through holes in cold screens and located approximately 10\,cm away from the sample. The PBr$_3$ adsorption was carried out at 77\,K with a partial pressure of $5 \cdot 10^{-10}$\,Torr for two minutes. After each experiment, the sample was heated at 1170\,K  for several hours to remove phosphorous from the surface from the previous experiment \cite{2005Brown}. We used mechanically cut Pt-Ir tips. The voltage ($U_s$) was applied to the sample. The WSXM software \cite{WSXM} was used to process all STM images.

\subsection{Computational methods}

The spin-polarized DFT calculations were performed using the Vienna \textit{ab initio} simulation package (VASP) \cite{1993Kresse, 1996Kresse}. The framework of Perdew-Burke-Ernzerhof (PBE) generalized gradient approximation (GGA) for the exchange-correlation potential \cite{1996Perdew}  was utilized. The projector augmented wave approach \cite{1999Kresse, 1994Blochl} with a kinetic energy cutoff of 350\,eV was used to simulate the interaction between ions and electrons. A slab consisting of eight Si layers with periodic 4$\times$4 supercells was used to represent the Si(100)-2$\times$1 surface. Adatoms were put on the upper surface, the hydrogens covered the bottom surface. During the optimization process, the bottom two Si layers were fixed in their bulk positions, while the coordinates of the remaining atoms were relaxed. The residual forces acting on the relaxed atoms were less than 0.01\,eV/\,{\AA}. We used a vacuum interval between slabs of 13.5\,{\AA} thickness to prevent the surface-surface interactions. The $\Gamma$-centered 4$\times$4$\times$1 k-point mesh was employed. STM images were calculated within the Tersoff-Hamann approximation \cite{1985Tersoff} at a voltage of $+2.5$\,V.

\section{Results}

\subsection{Experimental Results}

\begin{figure}[h]
\begin{center}
\includegraphics[width=\linewidth]{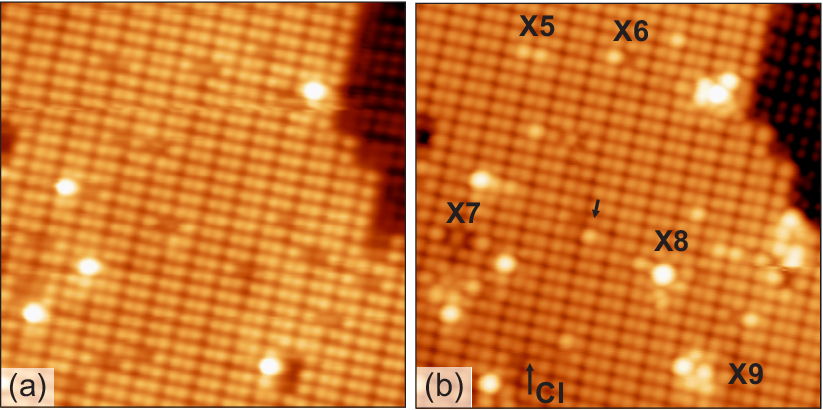}
\caption{\label{fig_S2} Chlorinated Si(100)-2$\times$1 surface before and after exposure to PBr$_3$. (a) Empty state STM image ($U_s =3.5$\,V, I$_t$ = 2.0\,nA) of the initial Si(100)-2$\times$1 surface covered with an incomplete monolayer of chlorine with a predominance of mono- and bivacancies. (b) Empty state STM image ($U_s =2.5$\,V, I$_t$ = 2.0\,nA) of the same surface area exposed by PBr$_3$. The objects formed in the vacancies are labelled as X5--X9. The arrow indicates a shift of Cl monovacancy to an adjacent silicon atom. The arrow with Cl denotes healing of the vacancy with chlorine.}
\end{center}
\end{figure}

Figure~\ref{fig_big}a shows a Si(100)-2$\times$1 surface covered with an incomplete monolayer of chlorine. Mono- and bivacancies predominate on the surface, with a small number of larger vacancies. Note that with these scanning parameters of the halogenated surface, the dark line represents the middle of the dimer row rather than the trough. Therefore, the bivacancy in the empty state STM image appears as a double protrusion centered on the dark line. In addition to chlorine, the surface contains low concentrations of iodine and very low concentrations of bromine (see section Experimental and computational details). Figure~\ref{fig_big}b shows the same surface area as in Fig.~\ref{fig_big}a after PBr$_3$ adsorption. Figure~\ref{fig_S2} shows another STM images of the surface area before and after PBr$_3$ adsorption. After exposure to PBr$_3$, most mono- and bivacancies were filled with molecular fragments. We labelled new objects, which in the vast majority of cases appeared after adsorption, as X1--X9.

In monovacancies, object X1 (Fig.~\ref{fig_big}b) was most frequently observed, which was a single protrusion with a height of 0.5\,{\AA} located above the Si atom. Very rarely we observed an X2 object with a double protrusion of height 1.2\,{\AA} located perpendicular to the row of dimers. The rest of the monovacancies remained unfilled.

In bivacancies we found a greater variety of structures, although some of them also remained empty. We mostly encountered object X3 (Fig.~\ref{fig_big}b), which was similar to object X2. We sometimes observed object X4, which has two protrusions in the bivacancy, identical  to objects X1, and a protrusion in the middle of the silicon dimer with a height of 0.1--0.2\,{\AA} below the Cl monolayer. Object X5 (Fig.~\ref{fig_S2}b) is formed by two objects X1, and object X6 consists of one X1 and one unfilled vacancy. In objects X7 and X8, the protrusions are located in the middle of the dimer, and the Cl atoms of nearby dimers appear 0.1\,{\AA} higher. Such a protrusion in object X7 is low and positioned symmetrically in the middle of the dimer as in X4, whereas the protrusion in object X8 is located asymmetrically and has a height of 1.2 A. Rarely, vacancies larger than the bivacancies were found on the surface. Object X9 (Fig.~\ref{fig_S2}b), found in a three-dimer vacancy, consists of three protrusions above the Si atoms and one protrusion at the bridge position in a Si dimer. Thus, the most common objects observed in vacancies after PBr$_3$ adsorption consist of protrusions located on top of Si atoms and in the bridge positions, as well as a pair of protrusions located perpendicular to the dimer row.

\begin{figure}[h]
 \includegraphics[width=\linewidth]{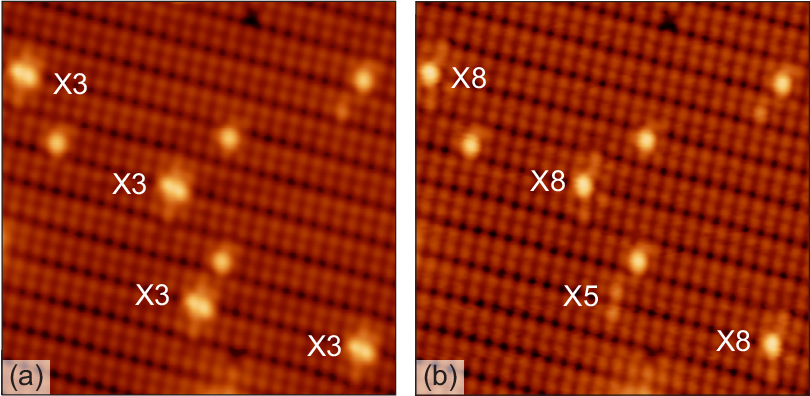}
\caption{\label{fig_tr} Transformation of object X3 into objects X5 and X8 during scanning. Empty state STM images (I$_t$ = 2.0\,nA, (a) $U_s =2.8$\,V, (b) $U_s =3.0$\,V) of the PBr$_3$-dosed Si(100)-2$\times$1 surface before (a) and after (b) transformation. }
\end{figure}

\begin{figure}[h]
 \includegraphics[width=\linewidth]{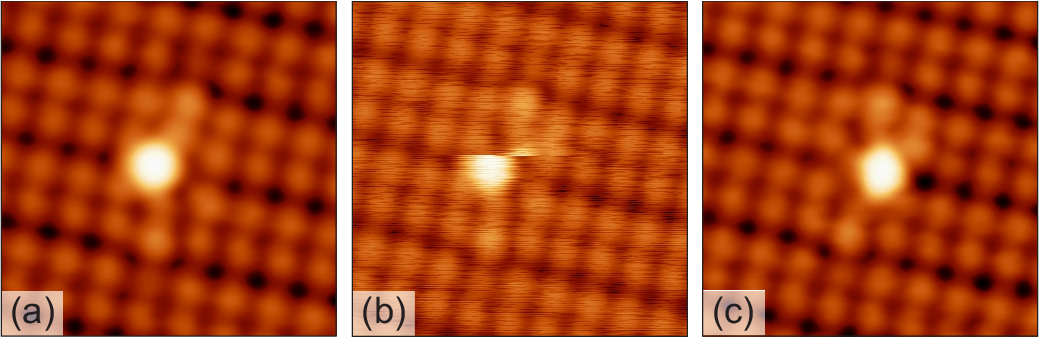}
\caption{\label{fig_dif} Switching between the two symmetrical configurations of object X8 while scanning. (a--c) Sequentially recorded empty state STM images ($U_s =2.8$\,V, I$_t$ = 2.0\,nA) of the Si(100)-2$\times$1 surface with object X8. The slow scan direction proceeded from bottom to top.}
\end{figure}

During the scanning process, we sometimes observed changes in objects. In particular, we observed the transformation of object X3 into objects X5 and X8 (Fig.~\ref{fig_tr}). Also, when scanning object X8, it switches between two configurations (Fig.~\ref{fig_dif}). For two different configurations of X8, the protrusion is located on opposite sides of the centerline of the Si dimer.

\begin{figure}[h]
\begin{center}
\includegraphics[width=\linewidth]{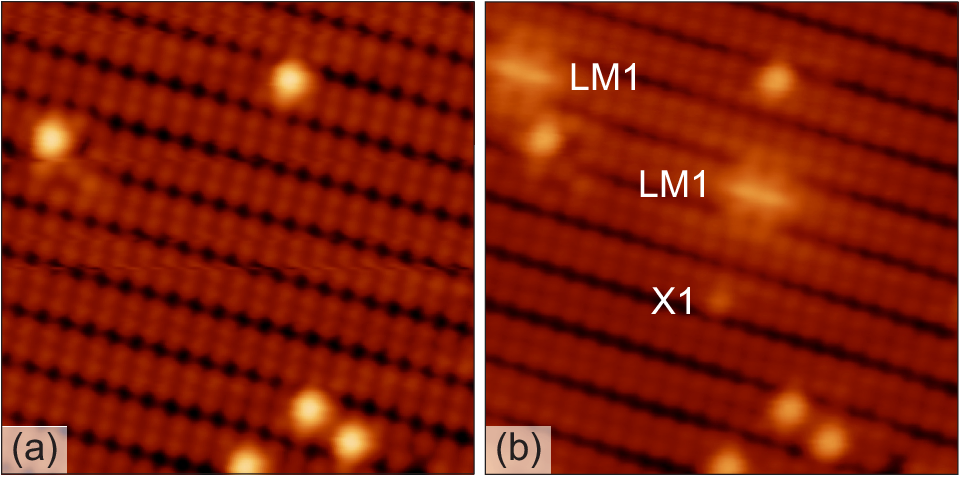}
\caption{\label{fig_S3} Empty state STM images ($U_s =2.5$\,V, I$_t$ = 2.0\,nA) of the Si(100)-2$\times$1-Cl surface without vacancies before (a) and after (b) exposure to PBr$_3$. A new object (X1) appeared on a defect-free surface area. Two objects LM1 also formed, which are halogen atoms in the groove between the dimer rows.}
\end{center}
\end{figure}

For the most part of the defect-free surface, no new objects appeared after adsorption, indicating that molecular fragments do not penetrate under the chlorine monolayer. Nevertheless, very rarely new objects appeared in the chlorine monolayer, as for example shown by the asterisk in Fig.~\ref{fig_big}b. Another example of Cl monolayer modification after PBr$_3$ adsorption is shown in Fig.~\ref{fig_S3}. There were no vacancies in the original surface area (Fig.~\ref{fig_S3}a), but after exposing the Cl monolayer to PBr$_3$ (Fig.~\ref{fig_S3}b), a new object appeared which looks like X1 in Fig.~\ref{fig_big}b. In addition to the appearance of new objects, we also sometimes observed a change in the position of vacancies, as shown by the arrow in Fig.~\ref{fig_big}b and in Fig.~\ref{fig_S2}b. Sometimes monovacancies were repaired by Cl atoms (arrow with Cl in Fig.~\ref{fig_S2}b).

\subsection{Computational Results}

\subsubsection{PBr$_3$ interaction with mono- and bivacancies on Si(100)-2$\times$1-Cl}

\begin{figure*}[t]
 \includegraphics[width=\linewidth]{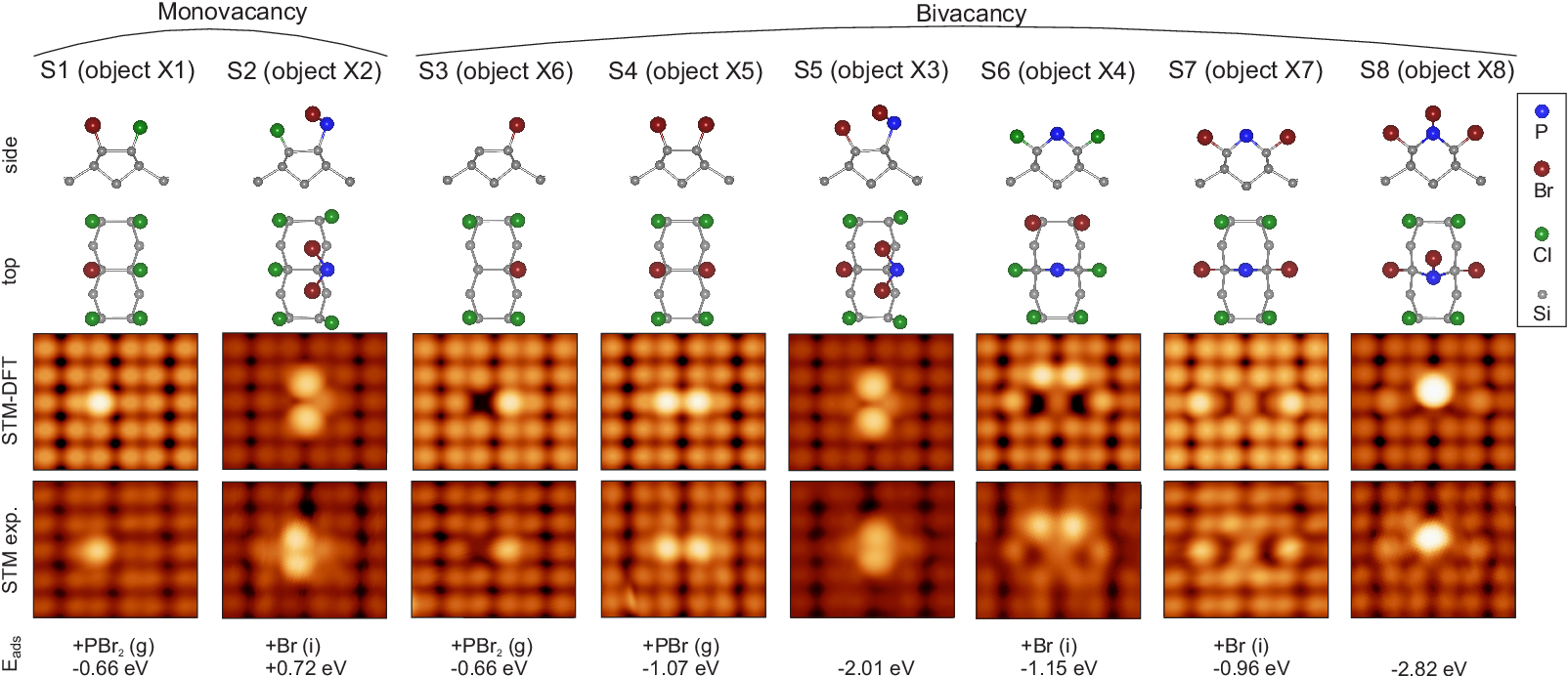}
\caption{\label{fig_str}
Calculated structures of PBr$_3$ fragments in mono- and bivacancies on the Si(100)-2$\times$1-Cl surface, corresponding to the observed objects X1--X8. For each structure S1--S8, side and top views, simulated empty state STM images, and experimental atomic-resolution empty state STM images are shown. The adsorption energies for each surface structure are listed below. For those structures in which not all atoms of the PBr$_3$ molecule are located in the vacancy, the adsorption energies are calculated with the remaining PBr$_2$ or PBr fragments in the gas phase 5\,{\AA} above the surface, and a single bromine atom Br(i) is placed in the groove between four dimers.}
\end{figure*}

To identify objects X1--X8, we carried out DFT modeling of various fragments of the PBr$_3$ molecule in mono- and bivacancies (Fig.~\ref{fig_str}). Our previous study of the completely dissociated molecule on the Si(100) surface revealed that Br atoms occupy positions above Si atoms, whereas P atoms are favored to be either in a groove or in a bridge position in the dimer \cite{2023Shevlyuga}. Taking this into account, we placed a Br atom on top of a Si atom in a Cl vacancy and found that the simulated STM image from such a structure (S1) coincides with the STM image of object X1, and the height in both cases is 0.5\,{\AA}. Thus, when approaching a Cl vacancy, the PBr$_3$ molecule dissociates, leaving a Br atom in the vacancy. When calculating the PBr$_3$ adsorption energy into a vacancy, we placed the PBr$_2$ fragment in a vacuum.

To determine the higher object X2 in a monovacancy, we calculated two structures with PBr$_3$ and PBr$_2$ in the vacancy. Unlike PBr$_3$, the simulated STM image from PBr$_2$ (structure S2) agrees well with the experimental one. To calculate the PBr$_3$ adsorption energy in a monovacancy with the formation of a PBr$_2$ fragment, we placed the remaining Br atom on the surface in the groove between the dimers (structure LM1 in Ref. \cite{2020PavlovaPRB}). The resulting adsorption energy of S2 is positive, which might be related to the less favorable site of the remaining Br under the chlorine monolayer. To explain the presence of X2 despite positive adsorption energy, we can suggest that either the remaining Br atom has moved to a more favorable site, or the chlorine monolayer has changed, resulting in a bivacancy instead of the monovacancy.

In bivacancies, we observed objects X5 and X6, consisting of one or two objects X1, which is a Br atom. Therefore, we modeled one and two Br atoms in a bivacancy (structures S3 and S4). The adsorption of two Br atoms with the removal of PBr into vacuum is energetically preferable to the adsorption of one Br atom with the removal of PBr$_2$ into the gas phase. We also calculated a surface structure with P and Br atoms on one Si dimer, each on top of a Si atom. In the simulated STM images, the P atom on top of the Si atom appears to be similar to the Br atom, making the STM image of such a structure looks like the STM image of structure S4. However, because the adsorption energy of such a system with the Br$_2$ removal into vacuum is positive, we believe that the incorporation of P and one Br atom is less likely than the incorporation of two Br atoms.

Object X3 in a bivacancy is very similar to object X2 in a monovacancy, so we also modeled it as a PBr$_2$ fragment on a Si atom, and placed the remaining Br atom in an adjacent vacancy (structure S5). Thus, object X3 is formed as a result of the PBr$_3$ dissociation into PBr$_2$ and Br on one Si dimer. This is the most favorable configuration, having a PBr$_2$ fragment on a clean Si(100) surface after PBr$_3$ dissociation \cite{2023Shevlyuga}. This dissociation pathway on a single Si dimer also occurs for other similar molecules, such as PH$_3$ \cite{2016Warschkow, 2005Warschkow}, PCl$_3$ \cite{2021Pavlova}, AsH$_3$ and NH$_3$ \cite{2001Miotto}.

The remaining objects X4, X7, and X8 contain a protrusion in the center of the dimer, which can be modeled by a P atom since the bridge position for it is one of the most favorable, unlike halogens \cite{2023Shevlyuga}. In the simulated STM image of structures S6 and S7 with a P atom in the bridge position, the protrusion has a low height, which is consistent with the experimental STM images of objects X4 and X7. Structures S6 and S7 differ in that the P atom is in a bridge position in the chlorinated and brominated Si dimer, respectively, and two Br atoms are located in the bivacancy in S6. Note that Cl (or Br) atoms can also occupy the bridge position in the chlorinated Si dimer \cite{2020PavlovaPRB}. However, such objects are mobile at a voltage of about +2.5 V, are usually surrounded by a bright halo because they are positively charged, and they usually transform into another object when the polarity changes \cite{2020PavlovaPRB}. Since objects X4 and X7 did not exhibit this behavior, we believe that the P atom, rather than the halogen atom, is in the bridge position.

Object X8 has a high protrusion, so we added a Br atom on top of the P atom. As a result of optimization of structure S8, the Br and P atoms were shifted from the middle of the dimer in opposite directions. Structure S8 has one more equivalent configuration, symmetrically reflected relative to the Si dimer, in which the P atom is displaced to one side of the dimer, and the Br atom to the other. During scanning, we observed a switch from one configuration to another (Fig.~\ref{fig_dif}).

Additionally, we observed object X9, formed in a vacancy from three neighboring Si dimers (Fig.~\ref{fig_S2}b). The STM image of X9 is similar to the STM image of the p-5 structure of Ref.~\cite{2023Shevlyuga}, which is a completely dissociated molecule on a clean silicon surface. In this structure, three Br atoms are located on top of the Si atoms, and a P atom is in a bridge position on the central Si dimer.

\subsubsection{PBr$_3$ interaction with a defect-free Si(100)-2$\times$1-Cl surface}

\begin{figure*}[t]
\begin{center}
\includegraphics[width=\linewidth]{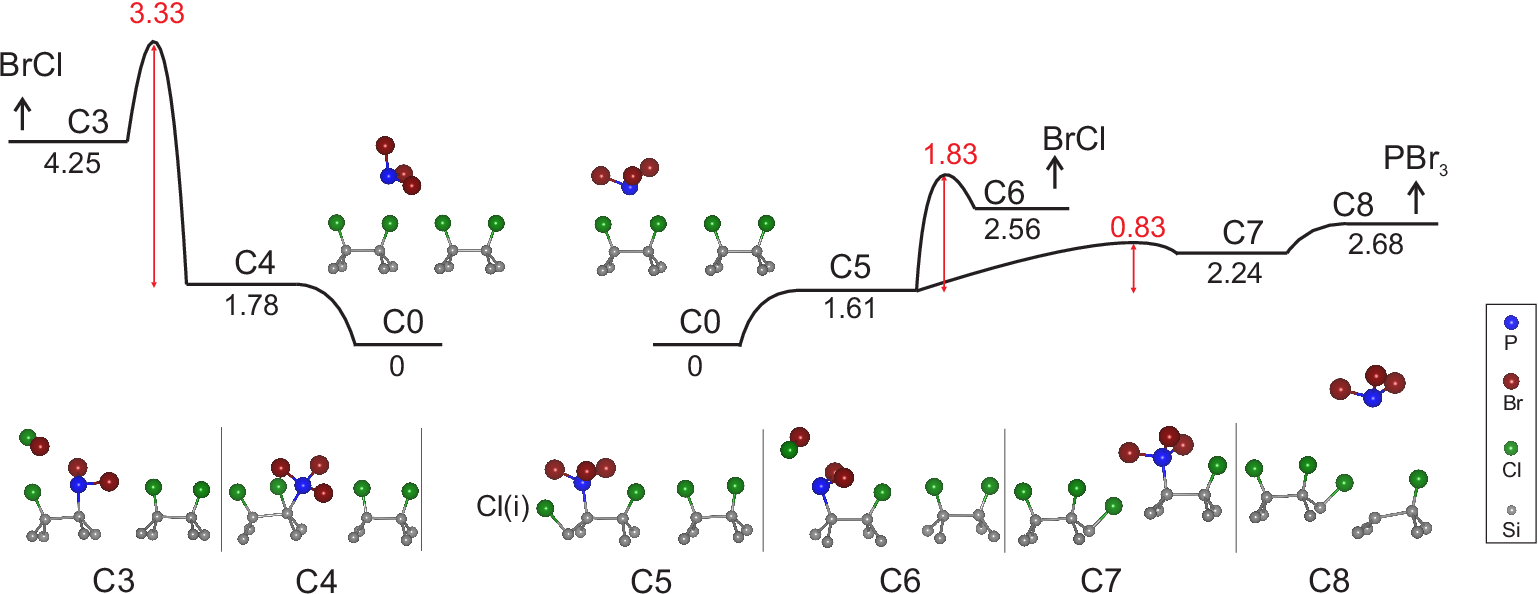}
\caption{\label{fig_s4} Energy barrier diagrams of PBr$_3$ adsorption on the Si(100)-2$\times$1-Cl surface. Pathways start at the center from the initial state C0. A side view is shown for each surface structure C0, C3--C8. All energies are given in eV relative to the initial state energy. Red numbers indicate activation barriers.}
\end{center}
\end{figure*}

To confirm the resistance of a chlorine monolayer to PBr$_3$ adsorption, we calculated the activation barriers to undesired molecule incorporation into a defect-free monolayer. Calculations were performed by analogy with Ref.~\cite{2021Pavlova} using the same parameters to compare activation barriers with those for other molecules and resists. The calculated configurations and activation barriers are shown in Fig.~\ref{fig_s4}. The molecule can form a P-Si bond by the displacement of a Cl atom with an activation barrier of 1.61 eV (structure C5). Diffusion of a displaced Cl atom has a barrier of 0.83 eV and does not lead to a decrease in the energy of the system (structure C7). PBr$_3$ can be easily removed from such a structure with a desorption energy of 0.44 eV (structure C8), which means that the P-Si bond will be broken and phosphorus will not be incorporated into silicon. The incorporation of the PBr$_2$ fragment is possible with the Br desorption in vacuum in the form of a BrCl molecule. The minimum barrier for the incorporation of a PBr$_2$ fragment was found to be 3.44\,eV (structure C6), which is comparable to the barrier for the PCl$_2$ incorporation under a chlorine monolayer (3.71\,eV) and significantly higher than the barrier for the PH$_2$ incorporation under a hydrogen monolayer (1.86\,eV) \cite{2021Pavlova}. Such a barrier (3.44 eV) is hard to overcome at 77 K (and even at room temperature), therefore, the incorporation of phosphorus under the Cl monolayer on Si(100)-2$\times$1 is very difficult.

\section{Discussion}

We were able to reveal the change in the perfect chlorine layer on the PBr$_3$-dosed surface because we scanned the same surface areas before and after PBr$_3$ adsorption. We observed only small modifications in the perfect chlorine monolayer after PBr$_3$ adsorption. In particular, we noticed the replacement of Cl with object X1 (Fig.~\ref{fig_S3}), which is a Br atom. The PBr$_3$ adsorption was carried out at 77 K, so Cl atoms should not be desorbed creating a monovacancy. We can explain the appearance of Br if we assume that after the PBr$_3$ dissociation in vacancies, some Br atoms insert into the Cl monolayer in the groove between the dimer rows (object LM1 in Ref.~\cite{2020PavlovaPRB}) or in the bridge position (object LM2 in Ref.~\cite{2020PavlovaPRB}). The introduced Br atoms can move like Cl atoms within the monolayer \cite{2020PavlovaPRB} and in some place displace a Cl atom. On the same surface area, we also observed two inserted atoms in the groove between the dimer rows (objects LM1 in Fig.~\ref{fig_S3}), one of which could be a Cl atom displaced by this Br atom. We would like to emphasize that in this case the resist is preserved, since one halogen atom is replaced by another, and the phosphorus incorporation does not occur.

We also observed a shift in the position of vacancies (arrows in Fig.~\ref{fig_big}b and Fig.~\ref{fig_S2}b). The vacancy displacement can be caused by stress in the Cl layer due to the Br insertion after the PBr$_3$ dissociation. We also very rarely encountered other inserted objects that we were unable to identify (as for example shown by the asterisk in Fig.~\ref{fig_big}b) and therefore cannot claim that phosphorus atoms are never embedded in a defect-free surface. Thus, we conclude that chlorine is a good resist for PBr$_3$ based on the very small number of penetration into defect-free areas of the monolayer, some of which are Br atoms that do not damage the resist. The calculations also confirm that a defect-free chlorine monolayer effectively protects silicon from PBr$_3$ penetration.

Using DFT calculations, we identified two structures that were commonly found in monovacancies (X1, X2) and the six most frequently observed structures in bivacancies (X3--X8) (Fig.~\ref{fig_str}). Monovacancies were predominantly filled with Br or remained unfilled. The PBr$_2$ fragment was observed very rarely, so in the vast majority of cases, P atoms do not form a bond with a silicon atoms in a monovacancies. Instead, the P atom formed bonds with the Si atom in bivacancies. In bivacancies, phosphorus either formed a bond with one Si atom (object X3); in other cases, P was in a bridge position in the dimer, forming bonds with two Si atoms (objects X4, X7, and X8). In addition, the bivacancies could remain unfilled, or one or two Br atoms could be observed in them (objects X5 and X6). In a vacancy of six Cl atoms, a completely dissociated molecule with the P atom in a bridge position and the Br atoms in on top sites was observed (object X9 in Fig.~\ref{fig_S2}b). This observation confirms that three adjacent silicon dimers are sufficient for complete dissociation of PBr$_3$.

During scanning, we observed the transformation of object X3 into objects X5 and X8. When X3 is transformed into X5, phosphorus is desorbed as PBr, and two Br atoms remain in the bivacancy. When X3 is transformed to X8, PBr$_2$ also dissociates to PBr, however PBr does not desorb but instead moves to the bridge position. Object X8 is more energetically favorable than X3, so we can expect to find scanning conditions that make X3 dissociation into X8 during scanning very likely.

Phosphorus can be introduced into silicon by dissociating of a phosphorus-containing fragment of the molecule by an STM pulse, as was reported for phosphine \cite{2022Wyrick}. Hydrogen bond breaking occurs with atomic accuracy by multi-electron vibrational heating due to the extra long lifetime of the stretching mode \cite{1995Shen}. However, unlike hydrogen, the removal of chlorine and bromine by an STM pulse can not be realized by an efficient vibrational mechanism since such a long-lived mode does not exist \cite{2020Pavlova, 2022Pavlova}. Here we demonstrated dissociation of the molecular fragment during scanning, which is less aggressive than dissociation by pulses (including for the STM tip), at least for halogens.

In all objects with a phosphorus atom that we were able to identify, the P atom is in the bridge position or P can shift into it during scanning. The P atom in the bridge position seems interesting for phosphorus incorporation into silicon, since it is already in the site of the next layer of the silicon crystal lattice. Therefore, P will be a substitutional impurity in silicon if the epitaxy process can be optimized so that there will be no diffusion of phosphorus. In contrast, the present approach of P placement by PH$_3$ adsorption requires complete dissociation of PH$_3$ to a single P atom in an end-bridge site in the row between Si dimers, followed by annealing to exchange the P atom with one of the neighboring Si atoms \cite{2012Fuechsle, 2022Wyrick}. In this case, an inaccuracy arises in the placement of P in the Si lattice since it is unpredictable which of the neighboring Si atoms the P atom will exchange.

\section{Conclusions}

The PBr$_3$ adsorption on the Si(100) surface covered with a chlorine monolayer with mono- and bivacancies was investigated. Since PBr$_3$ was adsorbed in the STM chamber, we were able to identify the adsorption sites and observe minor changes in the chlorine layer. Fragments of PBr$_3$ molecules were unable to penetrate through chlorine on most of the defect-free surface area, and calculations support the chlorine layer's strong resistance to PBr$_3$. We determined the positions of all the atoms of the molecule after dissociation in mono- and bivacancies. A phosphorus atom almost never forms a bond with silicon in a monovacancy. In a bivacancy, either a PBr$_2$ fragment is located on top of a Si atom, or PBr or P in a bridge position on the Si dimer. The possibility of PBr$_2$ dissociation onto PBr during scanning with phosphorus transfer to the bridge position has been demonstrated. In the bridge position, phosphorus is already in place of the atom of the next layer of the silicon crystal lattice. The results provide insight on what occurs when molecules like PBr$_3$ adsorb onto a monolayer containing mono- and bivacancies. These findings can be used to develop atomically precise methods for introducing impurities into silicon.

\section*{Acknowledgments}
This study was supported by the Russian Science Foundation under grant No. 21-12-00299. We also thank the Joint Supercomputer Center of RAS for providing the computing power.

\bibliography{PBr3_Si-Cl_arxiv}
\end{document}